\documentclass[traditabstract]{aa}  
\usepackage{savesymbol}
\usepackage{amsmath}
\savesymbol{iint}
\usepackage{txfonts}
\restoresymbol{TXF}{iint}
\usepackage{txfonts}
\usepackage{float}
\usepackage{placeins}
\usepackage{lscape}
\usepackage{natbib}
\usepackage{ulem}

\bibpunct{(}{)}{;}{a}{}{,}
\usepackage{graphicx}
\DeclareGraphicsExtensions{.eps}

\renewcommand{\deg}{\hbox{$^\circ$}}
\newcommand{\hmsm}[4]{{#1}^{\rm h}{#2}^{\rm m}{#3}^{\rm s}\!\!.\:\!{#4}}
\newcommand{\dmsm}[4]{\, {#1}^{\circ}{#2}^{\prime}{#3}^{\prime\prime}\!\!\!.\;\!{#4}}

\begin{document}
   \title{Revealing Hanny's Voorwerp: radio observations of IC~2497}

      \author{G. I. G. J\'ozsa\thanks{\email{jozsa@astron.nl}}
             \inst{1}
              \and
            M. A. Garrett
              \inst{1,2,3}
              \and
            T. A. Oosterloo
              \inst{1,4}
              \and
            H. Rampadarath
              \inst{2,5}
              \and
            Z. Paragi
              \inst{5,6}
              \and
            H. van Arkel
              \inst{1}
              \and
            C. Lintott
              \inst{7}
              \and
            W. C. Keel
              \inst{8}
              \and
            K. Schawinski
              \inst{9}
              \and
            E. Edmondson
              \inst{10}
      }
   \offprints{G. I. G. J\'ozsa}

   \institute{Netherlands Institute for Radio Astronomy, Postbus 2,
 7990 AA Dwingeloo, The Netherlands
\and
Leiden Observatory, Univ. Leiden, P.O. Box 9513 NL-2300 RA Leiden, The Netherlands
\and
Centre for Astrophysics and Supercomputing, Univ. Swinburne, Mail number H39, Swinburne University of Technology, PO Box 218, Hawthorn, Victoria 3122, Australia
\and
Kapteyn Astronomical Institute, Univ. Groningen, Postbus 800, 9700 AV Groningen, The Netherlands
\and
Joint Institute for VLBI in Europe, Postbus 2, 7990 AA Dwingeloo, The Netherlands
\and
MTA Research Group for Physical Geodesy and Geodynamics, P.O. Box 91, H-1521 Budapest, Hungary
\and 
Univ. Oxford, Dept. Physics, Denys Wilkinson Building, Keble Road, Oxford, OX1 3RH, UK
\and
Univ. Alabama, Dept. Physics \& Astronomy, Box 870324, University of Alabama, Tuscaloosa, AL 35487-0324, USA
\and
Univ. Yale, Dept. Physics, J.W. Gibbs Laboratory, 260 Whitney Avenue, Yale University, New Haven, CT 06511, USA
\and
Institute of Cosmology and Gravitation, Dennis Sciama Building, Burnaby Road, Portsmouth, PO1 3FX, UK
}
   \date{Received 29/04/2009 / Accepted 12/05/2009}

\abstract {
We present multi-wavelength radio observations in the direction of the
spiral galaxy IC~2497 and the neighbouring emission nebula known as
``Hanny's Voorwerp''.  Our WSRT continuum observations at $1.4\,\rm
GHz$ and $4.9\,\rm GHz$, reveal the presence of extended emission at
the position of the nebulosity, although the bulk of the emission remains
unresolved at the centre of the galaxy. e-VLBI $1.65\,\rm GHz$
observations show that on the milliarcsecond-scale a faint central
compact source is present in IC~2497 with a brightness temperature in
excess of $4\cdot 10^5\,\rm K$. With the WSRT, we detect a large
reservoir of neutral hydrogen in the proximity of IC~2497. One cloud
complex with a total mass of $5.6\cdot 10^9\,{M}_\odot$ to the South
of IC~2497, encompasses Hanny's Voorwerp. Another cloud complex is
located at the position of a small galaxy group $\sim 100\,\rm kpc$ to
the West of IC 2497 with a mass of $2.9\cdot\,10^9\,{M}_\odot$. Our
data hint at a physical connection between both complexes. We also
detect \ion{H}{i} in absorption against the central continuum source
of IC~2497.

Our observations strongly support the hypothesis that Hanny's Voorwerp
is being ionised by an AGN in the centre of IC~2497. In this scenario,
a plasma jet associated with the AGN, clears a path through the
ISM/IGM in the direction towards the nebulosity. The large-scale radio
continuum emission possibly originates from the interaction between
this jet and the large cloud complex that Hanny's Voorwerp is embedded
in. The \ion{H}{i} kinematics do not fit regular rotation, thus the
cloud complex around IC~2497 is probably of tidal origin. From the
\ion{H}{i} absorption against the central source, we derive a lower
limit of $ 2.8 \,\pm\, 0.4\cdot 10^{21}\, {\rm atoms}\, {\rm cm}^{-2}$
to the \ion{H}{i} column density. However, assuming non-standard conditions for
the detected gas, we
cannot exclude the possibility that the AGN in the centre of IC~2497
is Compton-thick.
}

\keywords{Galaxies: active, Galaxies: IGM, Galaxies: individual: IC~2497}
   \maketitle
%
\section{Introduction}
\label{Sect_01}

Dutch school teacher, Hanny van Arkel, discovered what is surely one
of the most bizarre objects uncovered via the GalaxyZoo.org
morphological census \citep[][]{Lintott08},
SDSS~J094103.80+344334.2. This object, now known as ``Hanny's
Voorwerp'', appears as an irregular cloud located $15 - 25\,\rm kpc$
in projection from the massive disk galaxy IC~2497 (see grey scale
plots in Figs.~\ref{Fig_01} and \ref{Fig_02}) and has a redshift
matching with the galaxy ($V_{\rm sys}=15056 \,\pm\, 40\,\rm km\,{\rm
s}^{-1}$, taken from the NASA Extragalactic Database NED) to within
$300\,{\rm km}\,{\rm s}^{-1}$ \citep[][]{Lintott09}. The detected
[\ion{O}{III}] $\rm \lambda 5007$ emission, dominating the optical
appearance of the cloud, is distributed over an area of roughly
$15^{\prime\prime}\times 25^{\prime\prime}$, corresponding to $15\,\rm
kpc\,\times\, 25 \,\rm kpc$ at the distance of IC~2497 ($D_{\rm 3K} =
210 \,\pm\, 15\,\rm Mpc$). A WHT spectrum shows strong line emission,
with high-ionisation lines (He~II, [Ne~V]) coextensive with the
continuum \citep[][]{Lintott09}. The quiescent kinematics with line
widths of less than $100\,\rm km\,s^{-1}$ \citep[][]{Lintott09} and a
global velocity gradient of $\sim 100 \,\rm km\,s^{-1}$, make
ionisation from photons probable as the predominant ionisation process
rather than ionisation from shocks.  The data do not indicate the
presence of any ionising source in the immediate proximity of the
nebulosity.

The current leading hypothesis is that Hanny's Voorwerp is being
illuminated and heated by an AGN situated at the centre of
IC~2497. The phenomenon has been studied in observations of other
objects \citep[][]{Morganti91,Fosbury98,Yoshida02,Croft06} and the
hypothesis is supported by the optical observations
\citep[][]{Lintott09}, showing an emission spectrum of the nucleus of
IC~2497 comparable to a low-ionization nuclear emission-line region
(LINER) or the narrow-line region of a Seyfert galaxy \citep[see
also][and Fig.~3 therein]{Morganti91}.  The FIRST survey catalogue
lists a radio continuum source situated at the central position of
IC~2497 \citep{White97}. Hence, Hanny's Voorwerp appears to be a prime
example of AGN feedback processes ionising the surrounding IGM.

However, there is some puzzling evidence against this scenario. 
A non-detection in a short X-ray observation with the \textit{Swift}
satellite \citep[][]{Lintott09} implies that the hypothetical AGN is
Compton-thick towards the observer, but not towards Hanny's
Voorwerp. The alternative explanation is that in the short time span
of about 80000 years that the radiation needs to reach the nebula, the
AGN activity has been reduced by a large factor or has even ceased
altogether.

The aim of this letter is to present continuum and \ion{H}{i} radio
observations of IC~2497 and Hanny's Voorwerp, conducted by the
Westerbork Synthesis Radio Telescope (WSRT) and by the European VLBI
Network (EVN). Our continuum data show extended radio emission at the
position of Hanny's Voorwerp. At VLBI resolution a compact source at
the centre of the galaxy is detected. This clearly supports the
hypothesis that Hanny's Voorwerp is illuminated and ionised by an AGN
hosted by IC~2497. We demonstrate that Hanny's Voorwerp is embedded in
a large cloud complex of neutral hydrogen, possibly being the remnant
of an interaction of IC~2497 with a galaxy group located at a
projected angular distance of $100^{\prime\prime}$ from IC~2497.

We describe the observations and the data reduction
(Sect.~\ref{Sect_02}), the data (Sect.~\ref{Sect_03}) and discuss our
results (Sect.~\ref{Sect_04}).
%
%
\section{Observations and data reduction}
\label{Sect_02}
\begin{figure}
\centering
\resizebox{0.9\hsize}{!}{\includegraphics[angle=270]{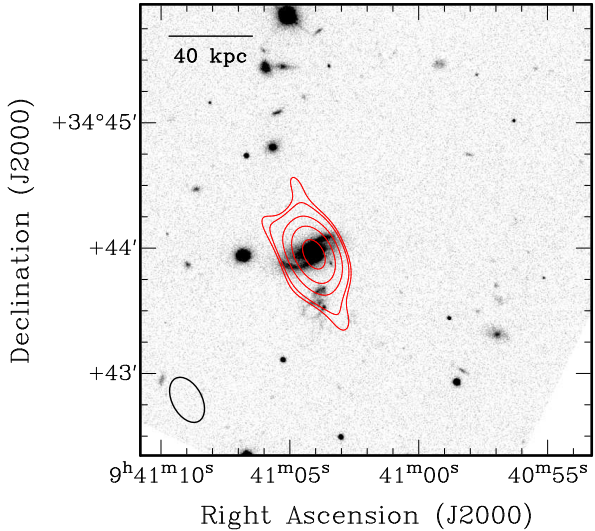}}
\caption{
WSRT 4.9-GHz continuum map (grey contours, red in online version)
overlaid on SDSS g-band image of IC~2497. Contours:
$0.6,0.75,1.5,3,6\,\rm mJy/beam$. The ellipse in the lower left corner
represents the clean beam HPBW (${\rm HPBW} =
23.3^{\prime\prime}\times 14.2^{\prime\prime}$, $\sigma_{\rm rms}=
0.15\, {\rm mJy}/{\rm beam}$). To the South of the galaxy, the optical
image shows Hanny's Voorwerp. }
\label{Fig_01}
\end{figure}

Radio continuum observations were performed with the WSRT as a service
project in two epochs on 28 September 2008 and 29 September 2008 with
a total integration time of 7.5 hours.  We alternated the receiver
frequency between $4.9\,\rm GHz$ and $1.4
\,\rm GHz$ with a total bandwidth of $160 \,\rm MHz$ to enable an
optimally uniform uv-coverage in each band.  The data underwent a
standard data reduction with the Australia Telescope National Facility
(ATNF) software package Miriad. The $4.9$-$\rm GHz$ map shown in
Fig.~\ref{Fig_01} was generated using visibilities with a baseline
length of $< 10\,{\rm k \rm \lambda}$ only and a robust weighting of 0.4. 
We found that this was the optimal weighting scheme to map the observed features (see Sect.~\ref{Sect_03}).
%
%
%
%

\begin{figure}
\centering
\resizebox{0.9\hsize}{!}{\includegraphics[angle=270]{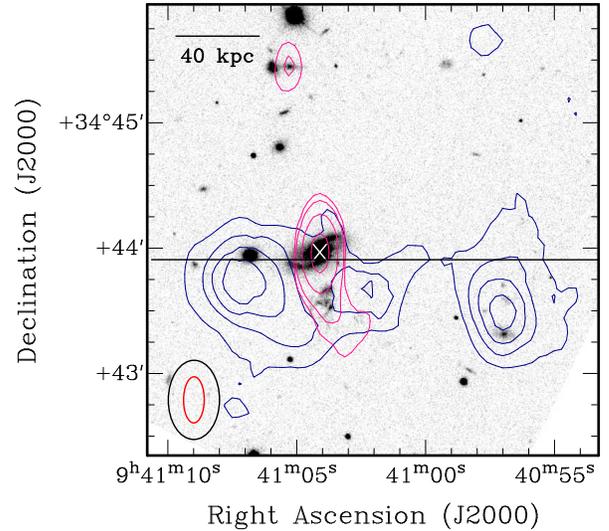}}
\caption{
\ion{H}{i} column density (dark grey, blue in online version,
contours: $0.5,1,1.5,2 \cdot\,10^{20}\,\rm atoms\,cm^{-2}$) and
$1.4$-$\rm GHz$ continuum map (light grey, pink in online version,
contours $0.35,0.7,2.8,11.2\,\rm mJy/beam$, $\sigma_{\rm rms} = 0.07
{\rm mJy}/{\rm beam}$) overlaid on SDSS g-band image of IC~2497 (see
Fig.~\ref{Fig_01}). The ellipses in the lower left corner represent
the clean beam HPBWs, the larger one for the \ion{H}{i} measurement
(${\rm HPBW} = 37.9^{\prime\prime}\times 24.5^{\prime\prime}$), the
smaller for the continuum image (${\rm HPBW} =
22.1^{\prime\prime}\times 9.9^{\prime\prime}$). The white cross marks
the position of the absorption spectrum shown in
Fig.~\ref{Fig_05}. The PV-diagram shown in Fig.~\ref{Fig_04} has been
taken along the solid line.}
\label{Fig_02}
\end{figure}

In addition, we observed IC~2497 in \ion{H}{i} with the WSRT for
$2\times 12 \rm h$ on 12 October 2008 and 28 October 2008.  We used a
total bandwidth of $20\,\rm MHz$, two parallel polarisations, and 1024
channels in total. After a standard data reduction with Miriad we
obtained a continuum map, and several data cubes each suited for the
specific analyses as described below. Due to the better spatial
resolution resulting from a complete uv-coverage, we use the
$1.4$-$\rm GHz$ continuum map from the \ion{H}{i} observations for our
analysis and the one from the dedicated continuum observations as a
consistency check. Both maps match within the errors.
Figure~\ref{Fig_02} shows both a total-\ion{H}{i} map, derived from a
data cube using natural weighting and a velocity resolution of
$108\,\rm km\,s^{-1}$, and the uniformly weighted $1.4$-$\rm GHz$
continuum map.
%
%
%
%

IC~2497 was observed by the European VLBI network (EVN) in phase
reference mode for 2 hours at $1.65\,\rm GHz$ on 30 September
2008. The observing bandwidth was 64 MHz in both LCP and RCP, with
2-bit sampling employed. The array included the Westerbork, Medicina,
Onsala 25-m, Torun, Effelsberg, Jodrell Bank MkII, and Darnhall
telescopes. The data from each telescope were transported to the
correlator at the Joint Institute for VLBI in Europe (JIVE) in
real-time, achieving a sustainable data rate of 512 Mbps.
For the observations, the target was phase-referenced to
J0945+3534, a VLBA calibrator located ~ 1.3 degrees away from the
target source (IC2497). 
The data analysis
with the National Radio Astronomical Observatory (NRAO) data reduction
package AIPS yielded a detection of a single compact source with a
signal-to-noise (SNR) $> 7$, located at RA
$\hmsm{09}{41}{04}{0875}\,\pm \, 0.\!\!^{\rm s}0002$ and Dec
$\dmsm{+34}{43}{57}{778}\,\pm\,0.\!\!^{\prime\prime}002$ (J2000) with a resolution of
$45\,{\rm milliarcseconds}\times 38\,{\rm milliarcseconds}$.
%
%
\section{Results}
\label{Sect_03}
While most of the continuum emission is unresolved in the WSRT
observations, positioned at the centre of the galaxy, it is evident
from the $1.4$-$\rm GHz$ map (Fig.~\ref{Fig_02}) that a faint
extension towards the SW is present, stretching out to the location of
Hanny's Voorwerp. Also at $4.9\,\rm GHz$ we detect extended radio emission
in addition to a dominating point source at the centre of
IC~2497 (Fig.~\ref{Fig_01}). 

\begin{figure}
\centering
\resizebox{0.9\hsize}{!}{\includegraphics{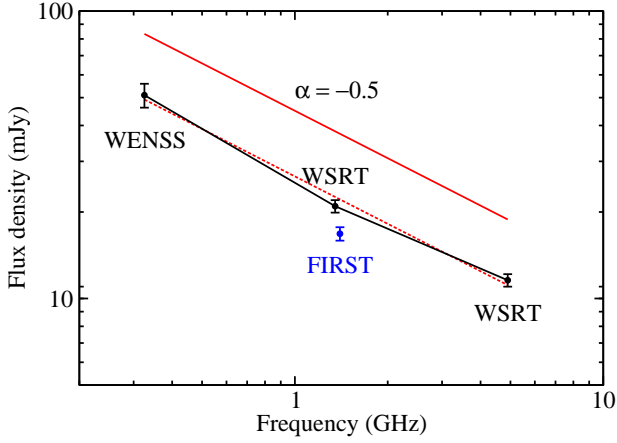}}
\caption{
Continuum flux densities as derived from the WSRT measurements. Also
shown is the FIRST flux density at $1.4\,\rm GHz$ and the WENSS result
at $0.325\,\rm GHz$. A power-law fit, represented by grey lines (red
in online version), yields a spectral index of $\alpha = -0.55$.}
\label{Fig_03}
\end{figure}

At $1.4 \,\rm GHz$ we measure a total flux density of $20.9 \,\pm\,
1.1\,\,\rm mJy$. This is confirmed in our second, independent
continuum map ($21.6\, \pm \,1.1\,\rm mJy$). The extended part of the
emission has a flux density of $3.2 \pm 0.2\,\rm mJy$. The FIRST
catalogue \citep{White97} reports a flux density of $16.8\,\pm\, 0.9
\,\rm mJy$, hence we conclude that probably the extended part of the emission
is not detected or resolved out in the FIRST snapshot observation.
Taking into account the flux density of $51.0\, \pm \,4.9\,\rm mJy$ measured at $325\,\rm MHz$
(Fig.~\ref{Fig_03}) as given in the Westerbork Northern Sky Survey
catalogue \citep[WENSS,][]{Rengelink97} and the flux densities of $20.9 \,\pm\,
1.1\,\,\rm mJy$ at $1.4\,\rm GHz$ and $11.6 \,\pm\,
0.6\,\,\rm mJy$ at $4.9\,\rm GHz$ as derived from our measurements, we fit a
power law with a spectral index of $-0.55\,\pm\, 0.05 $ (see
Fig.~\ref{Fig_03}).
%
%
%
%

We detect a central, unresolved source in IC~2497 in the WSRT
measurements at $1.4\,\rm GHz$ and $4.9\,\rm GHz$. The clear VLBI
detection shows the presence of a compact source at the centre of
IC~2497 with a flux density of $S_{\rm 1.65\,GHz, VLBI} = 1.09
\,\pm\, 0.14\,\rm mJy$. The observed position (RA
$\hmsm{09}{41}{04}{087}\,\pm \, 0.\!\!^{\rm s}0001$, Dec
$\dmsm{+34}{43}{57}{778}\,\pm\,0.\!\!^{\prime\prime}001$, J2000) is
offset approximately 230 milliarcseconds to the Southwest of the VLA
FIRST catalogue source at RA $\hmsm{09}{41}{04}{094}\,\pm
\, 0.\!\!^{\rm s}023$ and Dec
$\dmsm{+34}{43}{58}{00}\,\pm\,0.\!\!^{\prime\prime} 34$ (J2000,
errors at 90\% confidence level). Both positions are hence identical
within the errors.  An analysis of the VLBI measurement using the AIPS
routine IMFIT suggests a size of $<$ 60 milliarcseconds.  The flux
density ratio of about 1/20 between the central sources in VLBI- and
the WSRT measurements implies that within the WSRT beam the source
must be extended on intermediate scales.

\begin{figure}
\centering
\resizebox{0.9\hsize}{!}{\includegraphics[angle=270]{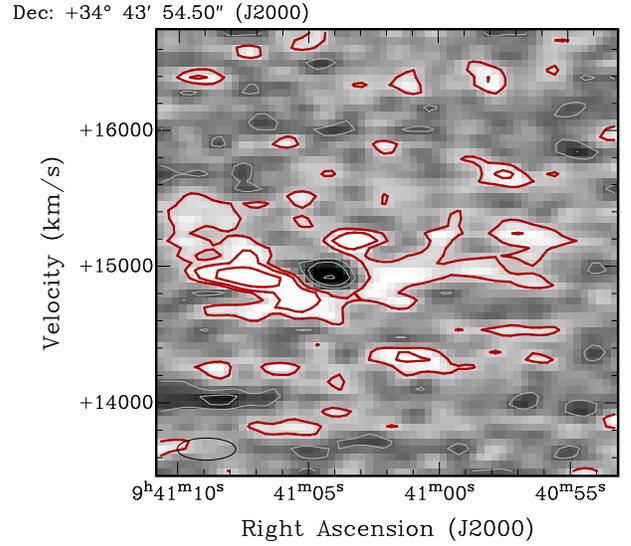}}
\caption{
PV-diagram taken along the black line shown in Fig~\ref{Fig_02}. The
Eastern and the Western cloud complexes are connected at a low
confidence level also in velocity-space. This indicates a physical
connection. The ellipse in the lower left corner indicates the
resolution (FWHM = $108\,\rm km\,s^{-1}$, minor axis HPBW =
$24.5^{\prime\prime}$). Contours: $-0.6, -0.3, -0.15\,\rm mJy/beam$
(light grey), $0.15, 0.3, 0.6\,\rm mJy/beam$ (dark grey, red in online
version), $\sigma_{\rm rms}= 0.1\, {\rm mJy}/{\rm beam}$.}
\label{Fig_04}
\end{figure}

We detect \ion{H}{i} in the vicinity of IC~2497 at velocities matching
the one of IC~2497 with a total mass of about $8.5\,\pm\, 2.1 \cdot
10^{9}\,M_\odot$ (see Fig.~\ref{Fig_02}). Due to the distance of
the observed objects, the detections are at a faint level, and we
assume an error of 20 percent in the column densities and the total
flux.
The neutral gas is concentrated in two cloud complexes. One, with a
mass of $M_{\rm {\ion{H}{i}}, E} = 5.6 \,\pm\, 1.4
\cdot\,10^9\,{M}_\odot$, is surrounding the Southern half of IC~2497
with a peak in column density of $2.5\cdot\,10^{20}\,\rm
atoms\,cm^{-2}$, roughly located at the position of a small galaxy to
the East of IC~2497. The other has a mass of $M_{\rm {\ion{H}{i}}, W}
= 2.9 \,\pm\, 0.8 \cdot\,10^9\,{M}_\odot$ and is located to the West
at a distance of $94^{\prime\prime}$ ($\hat{\sim} 96\, \rm kpc$) from
the centre of IC~2497, with a peak column-density of
$2.3\cdot\,10^{20}\,\rm atoms\,cm^{-2}$, located at the position of a
small galaxy group. At the position of IC~2497 itself we do not detect
any significant amount of \ion{H}{i} in emission. The kinematics of
the \ion{H}{i} is complex while showing a gradient towards higher
recession velocities from West to East (Fig.~\ref{Fig_04}). Apart from
the clear detection shown in Fig.~\ref{Fig_02}, our data hint at the
presence of gas spread out in the whole surroundings of IC~2497,
albeit at low significance. The most remarkable weak feature is a
connection between the Eastern and Western cloud complexes in space
and velocity, as indicated in the position-velocity (PV-) diagram in
Fig.~\ref{Fig_04}.
%
%
%
%
The Eastern cloud complex contains roughly 2/3 of the total detected
\ion{H}{i} and encompasses the location of Hanny's
Voorwerp. While we detect \ion{H}{i} at the position of the extended
continuum feature and the nebulosity with a redshift matching that of
Hanny's Voorwerp, the column density has a depression at that
position.

\begin{figure}
\centering
\resizebox{0.9\hsize}{!}{\includegraphics{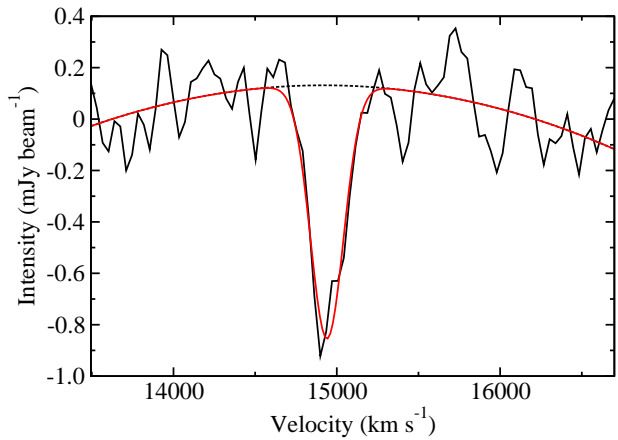}}
\caption{
\ion{H}{i} spectrum taken at the position indicated by the white cross in
Fig.~\ref{Fig_02}). In black the spectrum
is shown, the solid grey line (red in online version) is the result of
a Gaussian and second-order polynomial fit. The dashed curved line
shows the second-order baseline.
}
\label{Fig_05}
\end{figure}

\ion{H}{i} is also detected in absorption against the central
continuum source at the peak position of the
continuum. Fig.~\ref{Fig_05} shows the absorption feature as detected
in a robust-weighted data cube with a velocity resolution of $72\,\rm
km\,s^{-1}$.
%
%
%
%
The total absorbed \ion{H}{i} flux contained in a beam of $F^{\rm
abs}_{\rm \ion{H}{i}} = 0.28 \,\pm\, 0.03\,{\rm Jy}\,{\rm km}\,{\rm
s}^{-1}$ was determined from the spectrum, using a Gaussian- and
second-order polynomial fit for an additional background subtraction.
A Gaussian with a width of ${\rm FWHM}^{\rm abs}_{\rm \ion{H}{i}} =
226\,\pm\, 22 \, \rm km\,{\rm s}^{-1}$ (corrected for instrumental
broadening) represents the absorption profile well
(Fig.~\ref{Fig_05}). The \ion{H}{i} column density was estimated using
the flux density of the unresolved WSRT continuum detection at
$1.4\,\rm GHz$, under the assumption that the absorbing sheet covers
the background continuum source. We derive a column density of $N^{\rm
abs}_{\rm \ion{H}{i}} = 2.8 \,\pm\, 0.4\cdot 10^{21}\, {\rm atoms}\,
{\rm cm}^{-2}\,\frac{T_{\rm spin}}{100\,{\rm K}}\,\frac{1}{f}$, where
$T_{\rm spin}$ is the spin temperature and $f$ a covering factor of
the absorbing sheet with respect to the background source.
%
%
\section{Discussion}
\label{Sect_04}
Our observations support the hypothesis that IC~2497 contains an
active galactic nucleus with a radio jet emerging in the direction of
Hanny's Voorwerp.

From the flux density and the estimated size of the central continuum
source detected in our VLBI experiment, we determine a limit on the
brightness temperature of the source $T_{\rm b} \gtrsim 4\cdot
10^5\,\rm K$. This suggests that the source is probably related to AGN
activity in the core of IC2497.  Furthermore, with the WSRT we detect
extended continuum emission containing a considerable fraction of the
total flux density at $1.4\,\rm GHz$. The extended emission originates
at the centre of IC~2497 and points in the direction of Hanny's
Voorwerp. The extension is close to perpendicular 
(at an angle of $76\deg$ to $87\deg$) to the optical disk of
IC~2497 and reaches a considerable distance from the galaxy
(Fig.~\ref{Fig_02}). Most likely, it represents a large-scale radio
jet.

The substantial $1.4\,\rm GHz$ radio luminosity $P_{1.4\,\rm GHz} =
1.1 \,\pm\, 0.2 \cdot 10^{23}\,\rm W\,Hz^{-1}$ and the spectral index
$\alpha = -0.55 \,\pm\, 0.05$ are consistent with that
picture. However, we note that this could also be associated with
strong star formation
\citep[e.g.][]{Sadler02}. 

The Eastern \ion{H}{i} complex has a depression in column density at
the position of Hanny's Voorwerp and the extended continuum
emission. 
It seems that the radiation from the central source
can reach the gas associated in Hanny's Voorwerp and ionise it. As a
consequence the density of the neutral gas is lower at the position of
the nebulosity.
The asymmetric radio jet appears to be visible only in connection to
the gas complex detected in \ion{H}{i}, where it passes through the
IGM, suggesting that it has cleared a path for the radiation from the
AGN to reach Hanny's Voorwerp.

Hanny's Voorwerp is thus not an isolated gas cloud. It is embedded in
and part of a very large gas complex spanning about $110\,\rm kpc$ in
projection. The location of the cloud complex and its irregular
morphology and kinematics, makes it rather likely that the cloud
complex is external in origin.  The presence of the Western cloud
complex at the location of a galaxy group with rather small members
and the tentative kinematical and morphological connection with the
Eastern cloud complex, suggests a scenario in which both complexes
belong to the same structure that contains gas stripped from the
galaxy group, instead of gas that has been stripped from a single
galaxy. Another candidate as a gas donator or another member of the
galaxy group would be the galaxy located to the East of IC~2497 close
to the peak of the total \ion{H}{i} intensity. In favour of a tidal
interaction with a more massive companion rather than a dwarf galaxy
is also the appearance of IC~2497 in the optical. IC~2497 seems to
exhibit a massive warp.

The detection of \ion{H}{i} in absorption already shows that the
radiation emerging from the centre of IC~2497 towards us passes
through the ISM/IGM at the location of IC~2497. 
The \ion{H}{i} column density of the absorbing material as given above
is a lower limit, assuming that the absorbing material is part of the
cool, extended ISM. This assumption is supported by
extended dust-lanes crossing the centre of the galaxy, already visible
in the SDSS images, indicating the existence of such an extended
component of the ISM \citep[see also][]{Lintott09}.  We can, however,
not exclude a scenario in which (part of) the absorption takes place
closer to the nucleus, in a compact, circumnuclear disk.  This might
scale the \ion{H}{i}-column density by two factors. Firstly, the
covering factor $f$ might differ substantially from 1. Scaling the
\ion{H}{i} column density by a factor of up to $1/f \gtrsim 10$ does
does not contradict our results, taking into account that we measure a
ratio of low- and high resolution flux densities of $\sim 20$.  The
second factor is the assumed spin temperature of the standard value of
$100\,\rm K$ for
\ion{H}{i} in normal spiral galaxies. 
Under the assumption of a covering factor of 0.1, a shielding column density of $\gtrsim 10^{24}\, {\rm atoms}\,{\rm cm}^{-2}$ would imply a spin-temperature of $\gtrsim 3600 K$, a realistic value for a gas residing in a circumnuclear disk
\citep[][]{Bahcall69}.
Hence, the \ion{H}{i} column density might
well reach the Compton-thick regime.
This would offer an alternative to a
short-time variability of the AGN as an explanation for non-detections
of the AGN at other wavelength regimes. Sensitive radio observations
at high resolution and forthcoming X-ray observations with \textit{Suzaku} and \textit{XMM-Newton} will help to solve
this issue.

In conclusion, our observations consistently support the picture that
the nebulosity called Hanny's Voorwerp is being illuminated and heated
by an AGN situated at the centre of IC~2497. The emission nebula is
part of a large cloud complex detected in \ion{H}{i}. We detect
\ion{H}{i} in absorption against the central continuum source,
indicating the possibility that the AGN is Compton-thick and hence not
yet detected at other wavelengths. Future observations will help to
clarify this.
%
%
\begin{acknowledgements}
The Westerbork Synthesis Radio Telescope (WSRT) is operated by the
ASTRON (Netherlands Foundation for Research in Astronomy) with support
from the Netherlands Foundation for Scientific Research (NWO). We
thank the director of the Radio Observatory division of ASTRON, Rene
Vermeulen, for allocating Director's time for this project, and the
WSRT staff for making the observations possible on short notice.\\
The European VLBI Network is a joint facility of European, Chinese,
South African and other radio astronomy institutes funded by their
national research councils. e-VLBI developments in Europe are
supported by the EC DG-INFSO funded Communications Network Development
project 'EXPReS', Contract No. 02662. We thank the EVN PC chair,
Tiziana Venturi, for her approval of the e-EVN observations on short
notice.\\
This research was supported by the EU Framework 6 Marie Curie Early
Stage Training programme under contract number MEST-CT-2005-19669
"ESTRELA".\\
This research has made use of the NASA/IPAC Extragalactic Database
(NED).\\
This work has made use of the data releases of the Sloan Digital Sky
Survey \citep[][]{Adelman-McCarhty08}.
\end{acknowledgements}
%
%
\bibliographystyle{aa}
\bibliography{hanny}
%
%
\end{document}